\documentclass[letterpaper,twocolumn,notitlepage,preprintnumbers]{quantumarticle}
\pdfoutput=1
\usepackage[utf8]{inputenc}
\usepackage[bookmarks=false,colorlinks=true,urlcolor=blue,citecolor=blue,linkcolor=blue]{hyperref}
\usepackage[numbers,sort&compress]{natbib}

\usepackage{amsmath,amssymb,mathtools}
\usepackage{graphicx}
\usepackage{bm,color}
\usepackage[]{cleveref}
\usepackage{multirow}
\usepackage{datetime}
\usepackage{mathtools}

\usepackage{graphicx}
\usepackage{dcolumn}
\usepackage{bm}
\usepackage{url}

\usepackage{algorithm}
\usepackage{algpseudocode}

\usepackage{color}
\usepackage{braket}
\usepackage{textcomp}

\begin{document}

\title{Quantum Logic Gate Synthesis as a Markov Decision Process}

\author{M. Sohaib Alam}
\email{sohaib.alam@nasa.gov}
\affiliation{Rigetti Computing, 2919 Seventh Street, Berkeley, CA, 94710-2704 USA}
\affiliation{Quantum Artificial Intelligence Laboratory (QuAIL), NASA Ames Research Center, Moffett Field, CA, 94035, USA}
\affiliation{USRA Research Institute for Advanced Computer Science (RIACS), Mountain View, CA, 94043, USA}

\author{Noah F. Berthusen}
\altaffiliation[Present address: ]{Department of Computer Science, University of Maryland, College Park, MD, 20742, USA}
\affiliation{Ames Laboratory, Ames, Iowa 50011, USA}
\affiliation{Department of Electrical and Computer Engineering, Iowa State University, Ames, Iowa 50011, USA}

\author{Peter P. Orth}
\affiliation{Ames Laboratory, Ames, Iowa 50011, USA}
\affiliation{Department of Physics and Astronomy, Iowa State University, Ames, Iowa 50011, USA}

\newdate{date}{2}{07}{2022}
\date{\displaydate{date}}

\begin{abstract}
Reinforcement learning has witnessed recent applications to a variety of tasks in quantum programming. The underlying assumption is that those tasks could be modeled as Markov Decision Processes (MDPs). Here, we investigate the feasibility of this assumption by exploring its consequences for two fundamental tasks in quantum programming: state preparation and gate compilation. By forming discrete MDPs, focusing exclusively on the single-qubit case (both with and without noise), we solve for the optimal policy exactly through policy iteration. We find optimal paths that correspond to the shortest possible sequence of gates to prepare a state, or compile a gate, up to some target accuracy. As an example, we find sequences of $H$ and $T$ gates with length as small as $11$ producing $\sim 99\%$ fidelity for states of the form $(HT)^{n} |0\rangle$ with values as large as $n=10^{10}$. In the presence of gate noise, we demonstrate how the optimal policy adapts to the effects of noisy gates in order to achieve a higher state fidelity. Our work shows that one can meaningfully impose a discrete, stochastic and Markovian nature to a continuous, deterministic and non-Markovian quantum evolution, and provides theoretical insight into why reinforcement learning may be successfully used to find optimally short gate sequences in quantum programming.

\end{abstract}

\maketitle

\section{Introduction}
Recent years have seen dramatic advances in the field of artificial intelligence \cite{Norvig-russell-ai} and machine learning \cite{Schwartz-David, Goodfellow-et-al-2016}. A long term goal is to create agents that can carry out complicated tasks in an autonomous manner, relatively free of human input. One of the approaches that has gained popularity in this regard is reinforcement learning. This could be thought of as referring to a rather broad set of techniques that aim to solve some task based on a reward-based mechanism \cite{Sutton-barto}. Formally, reinforcement learning models the interaction of an \textit{agent} with its \textit{environment} as a Markov Decision Process (MDP). In many practical situations, the agent may have limited access to the environment, whose dynamics can be quite complicated. In all such situations, the goal of reinforcement learning is to learn or estimate the \textit{optimal policy}, which specifies the (conditional) probabilities of performing actions given that the agent finds itself in some particular state.
On the other hand, in fairly simple environments such as the textbook grid-world scenario \cite{Sutton-barto}, the dynamics can be fairly simple to learn. Moreover, the state and action spaces are finite and small, allowing for simple tabular methods instead of more complicated methods that would, for example, necessitate the use of artificial neural networks \cite{Goodfellow-et-al-2016}. In particular, one could use the dynamic programming method of \textit{policy iteration} to solve for the optimal policy exactly \cite{Bellman}.

In recent times, reinforcement learning has met with success in a variety of quantum programming tasks, such as error correction \cite{RL-QEC}, combinatorial optimization problems \cite{RL-COP}, as well as state preparation \cite{Bukov-et-al, Bukov2-et-al, August-et-al, Arri-et-al, Zhang-et-al} and gate design \cite{An-et-al, Niu-et-al} in the context of noisy control. Here, we investigate the question of state preparation and gate compilation in the context of abstract logic gates, and ask whether reinforcement learning could be successfully applied to learn the optimal gate sequences to prepare some given quantum state, or compile a specified quantum gate. Instead of exploring the efficacy of any one particular reinforcement method, we investigate whether it is even feasible to model these tasks as MDPs. By discretizing state and action spaces in this context, we circumvent questions and challenges involving convergence rates, reward sparsity, and hyper-paremeter optimization that typically show up in reinforcement learning scenarios. Instead, the discretization allows us to exactly solve for and study quite explicitly the properties of the optimal policy itself. This allows us to test whether we can recover optimally short programs using reinforcement learning techniques in quantum programming situations where we already have well-established notions of what those optimally short programs, or circuits, should look like.

There have been numerous previous studies in the general problem of quantum compilation, including but not limited to, the Solovay-Kitaev algorithm \cite{Solovay-Kitaev-algo}, quantum Shannon decomposition \cite{Shende-et-al}, approximate compilation \cite{EricP-et-al, Vadym-et-al}, as well as optimal circuit synthesis \cite{Mosca-et-al, Vatan-et-al, Ross-et-al}. Here, we aim to show that optimally short circuits could be found through solving discrete MDPs, and that these circuits agree with independently calculated shortest possible gate sequences for the same tasks. Since the initial posting of this work, numerous works have continued to explore the interface between classical reinforcement learning and quantum computing. These include finding optimal parameters in variational quantum circuits \cite{pmlr-v107-yao20a,PhysRevResearch.2.033446,https://doi.org/10.48550/arxiv.2109.03188}, quantum versions of reinforcement learning and related methods \cite{https://doi.org/10.48550/arxiv.2103.05577,https://doi.org/10.48550/arxiv.2103.15084,Wang_You_Li_Childs_2021,https://doi.org/10.48550/arxiv.2011.11654,https://doi.org/10.48550/arxiv.2108.13050}, Bell tests \cite{PhysRevLett.125.160401}, as well as quantum control \cite{https://doi.org/10.48550/arxiv.2104.14539,RL_coherent_transport,https://doi.org/10.48550/arxiv.2101.09020,GIANNELLI2022128054}, state engineering and gate compilation \cite{rl4stateeng,https://doi.org/10.48550/arxiv.2007.14608,https://doi.org/10.48550/arxiv.2007.15957,PRXQuantum.2.010324,10.21468/SciPostPhysLectNotes.29,RL_compile_nature}, the subject of this paper.

In such studies, reinforcement learning is employed as an approximate solver of some underlying MDP. This raises the important question of how, and under what conditions, can the underlying MDP be solved exactly, and what kind of solution quality does it result in. Naturally, such MDPs can only be solved exactly for relatively small problem sizes. Our paper explores the answer to this question in the context of single-qubit state preparation and gate compilation, and demonstrates the effects of native gate choice, coordinate representation, discretization effects as well as noise.

The organization of this paper is as follows. We first briefly review the formalism of MDPs. We then investigate the problem of single-qubit state preparation using a discretized version of the continuous $\lbrace RZ, RY\rbrace$ gates, as well as the discrete gateset $\lbrace I, H, S, T\rbrace$. We then study this problem in the context of noisy quantum channels. Finally, we consider the application to the problem of single-qubit compilation into the $\lbrace H, T \rbrace$ gateset, and show, among other things, that learning the MDP can be highly sensitive to the choice of coordinates for the unitaries.

\subsection{Brief Review of MDPs}
Markov Decision Processes (MDPs) provide a convenient framing of problems involving an agent interacting with an environment. At discrete time steps $t$, an agent receives a representation of the environment's state $s_t \in \mathcal{S}$, takes an action $a_t \in \mathcal{A}$, and then receives a scalar reward $r_{t+1} \in \mathcal{R}$. The \textit{policy} of the agent, describing the conditional probability $\pi (a \vert s)$ of taking action $a$ given the state $s$, is independent of the environment's state at previous time steps and therefore satisfies the Markov property. The \textit{discounted return} that an agent receives from the environment after time step $t$ is defined as $G_t = \sum_{k=0}^{\infty} \gamma^{k} r_{t+k+1}$ where $\gamma \in [0, 1]$ is the \textit{discount factor}. The goal of the agent is then to find the optimal policy $\pi^{*} (a \vert s)$ that maximizes the \textit{state-value function} (henceforth, ``value function" for brevity), defined as the expectation value of the discounted return received from starting in state $s_t \in \mathcal{S}$ and thereafter following the policy $\pi (a \vert s)$, and expressed as $V_{\pi} (s) = \mathbb{E}_{\pi} \left[ G_t \vert s_t = s \right]$. More formally then, the optimal policy $\pi^{*}$ satisfies the inequality $V_{\pi^{*}} (s) \geq V_{\pi} (s)$ for all $s \in \mathcal{S}$ and all policies $\pi$. For finite MDPs, there always exists a deterministic optimal policy, which is not necessarily unique. The value function for the optimal policy is then defined as the \textit{optimal value function} $V^{*} (s) = V_{\pi^{*}} (s) = \text{max}_{\pi} V_{\pi} (s)$ for all $s \in \mathcal{S}$. 

The value function satisfies a recursive relationship known as the \textit{Bellman equation}
\begin{equation}
V_{\pi} (s) = \sum_{a} \pi (a \vert s) \sum_{s^{\prime}, r} p (s^{\prime}, r \vert s, a) \left[ r + \gamma V_{\pi} (s^{\prime}) \right]
\label{eq:bellman}
\end{equation}
relating the value of the current state to that of its possible successor states following the policy $\pi$. Note that the conditional probability of finding state $s^{\prime}$ and receiving reward $r$ having performed action $a$ in state $s$ specifies the environment dynamics, and also satisfies the Markov property. This equation can be turned into an iterative procedure known as \textit{iterative policy evaluation}
\begin{equation}
V_{k+1}(s) = \sum_{a} \pi (a \vert s) \sum_{s^{\prime}, r} p (s^{\prime}, r \vert s, a) \left[ r + \gamma V_{k} (s^{\prime}) \right]
\label{eq:bellman-iterate}
\end{equation}
which converges to the fixed point $V_{k} = V_{\pi}$ in the $k \rightarrow \infty$ limit, and can be used to obtain the value function corresponding to a given policy $\pi$. In practice, we define convergence as $\vert V_{k+1} - V_{k} \vert < \epsilon$ for some sufficiently small $\epsilon$.

Having found the value function, we could then ask if the policy that produced this value function could be further improved. To do so, we need the \textit{state-action value function} $Q_{\pi} (s, a)$, defined as the expected return by carrying out action $a$ in state $s$ and thereafter following the policy $\pi$, i.e. $Q_{\pi} (s, a) = \mathbb{E}_{\pi} \left[ G_t \vert s_t = s, a_t = a \right]$. According to the \textit{policy improvement theorem}, given deterministic policies $\pi$ and $\pi^{\prime}$, the inequality $Q_{\pi} (s, \pi^{\prime}(s)) \geq V_{\pi}(s)$ implies $V_{\pi^{\prime}}(s) \geq V_{\pi}(s)$ where $\pi^{\prime}(s) = a$ (and in general  $\pi^{\prime}(s) \neq \pi(s)$) for all $s \in \mathcal{S}$. In other words, having found the state-value function corresponding to some policy, we can then improve upon that policy by iterating through the action space $\mathcal{A}$ while maintaining the next-step state-value functions on the right hand side of Eq. (\ref{eq:bellman-iterate}) to find a better policy than the current one ($\epsilon$-greedy algorithm for policy improvement). 

We can then alternate between policy evaluation and policy improvement in a process known as \textit{policy iteration} to obtain the optimal policy~\cite{Sutton-barto}. Schematically, this process involves evaluating the value function for some given policy up to some small convergence factor, followed by the improvement of the policy that produced this value function. The process terminates when the improved policy stops differing from the policy in the previous iteration. Of course, this procedure to identify the optimal policy for an MDP relies on the finiteness of state and action spaces. As we will see below, by discretizing the space of 1-qubit states (i.e. the surface and interior of the Bloch sphere corresponding to pure and mixed states), as well as identifying a finite gate set, we create an MDP with the goal of state preparation for which optimal policies in the form of optimal (i.e. shortest) quantum circuits may be found through this method.

We note that one could view state evolution under unitary operations or left multiplication of unitaries by other unitaries as deterministic processes. These could be thought of as trivially forming a Markov Decision Process where the probabilities $p(s^{\prime} | s, a)$ have a $\delta$-function support on some (point-like) state $s^{\prime}$. Once we impose discretization, this underlying determinism implies that the dynamics of the discrete states are strictly speaking non-Markovian, i.e. the conditional probability of landing in some discrete state $s^{\prime}$ depends not just on the previous discrete state and action, but also on all the previous states and actions, since the underlying continuous/point-like state evolves deterministically. However, we shall see below that with sufficient care, both the tasks of state preparation and gate compilation can be modeled and solved as MDPs even with discretized state spaces. 

\section{Preparation of single-qubit states}
In this section, we will discuss the preparation of single-qubit states as an MDP. In particular, we will focus on preparing a discrete version of the $|1\rangle$ state. We will do so using two different gate sets, a discretized version of the continuous $RZ$ and $RY$ gates, and the set of naturally discrete gates $I$, $H$, $S$ and $T$, and describe probabilistic shuffling within discrete states to arrive at optimal quantum programs via optimal policies. We will also consider states of the form $(HT)^{n} |0\rangle$.

\subsection{State and Action Spaces}
We apply a fairly simple scheme for the discretization of the space of pure 1-qubit states. As is well known, this space has a one-to-one correspondence with points on a 2-sphere, commonly known as the Bloch sphere. With $\theta \in [0, \pi]$ denoting the polar angle and $\phi \in [0, 2 \pi)$ denoting the azimuthal angle, an arbitrary pure 1-qubit state can be represented as
\begin{equation}
\vert \psi \rangle = \text{cos} \left( \frac{\theta}{2} \right) \vert 0 \rangle + e^{i \phi} \text{sin} \left( \frac{\theta}{2} \right) \vert 1 \rangle
\end{equation}
The discretization we adopt here is as follows. First, we fix some small number $\epsilon = \pi / k$ for some positive integer $k$. Next, we identify polar caps around the north ($\theta = 0$) and south ($\theta = \pi$) pole. The northern polar cap is identified as the set of all 1-qubit (pure) states for which $\theta < \epsilon$ for some fixed $\epsilon$, regardless of the value of $\phi$. Similarly, the southern polar cap is identified as the set of all 1-qubit (pure) states for which $\theta > \pi - \epsilon$, independent of $\phi$. Apart from these special regions, the set of points $n \epsilon \leq \theta \leq (n+1) \epsilon$ and $m \epsilon \leq \phi \leq (m+1) \epsilon$ for some positive integers $1 \leq n \leq k-2$ and $0 \leq m \leq 2 k-1$ are identified as the same region. The polar caps thus correspond to $n = 0, k-1$, respectively. 

We identify every region $(n,m)$ as a ``state" in the MDP. As a result of this identification, elements of the space of 1-qubit pure states are mapped onto a discrete set of states such that the 1-qubit states can now only be identified up to some threshold fidelity. For instance, the $\vert 0 \rangle$ state is identified as the northern polar cap with fidelity $\text{cos}^{2} \left( \frac{\pi}{2k}\right)$. Similarly, the $\vert 1 \rangle$ state is identified with the southern polar cap with fidelity $\text{sin}^{2} \left( \frac{(k-1)\,\pi}{2k} \right) = \text{cos}^{2} \left( \frac{\pi}{2k}\right)$. In other words, if we were to try and obtain these states using this scheme, we would only be able to obtain them up to these fidelities.

Having identified a finite state space $\mathcal{S}$ composed of discrete regions of the Bloch sphere, we next identify single-qubit unitary operations, or gates, as the action space $\mathcal{A}$. There are some natural single-qubit gate sets that are already discrete, such as $\lbrace H, T \rbrace$. Others, such as the continuous rotation gates $\{RZ, RY\}$, require discretization similar to that of the continuous state space of the Bloch sphere. We discretize the continuous gates $RZ(\beta)$ and $RY(\gamma)$ by discretizing the angles $\beta, \gamma \in [0, 2 \pi]$. The resolution $\delta = \pi / l$ must be sufficiently smaller than that of the state space $\epsilon = \pi / k$ so that all states $s \in \mathcal{S}$ are accessible from all others via the discretized gateset $a \in \mathcal{A}$. In practice, a ratio of $\epsilon / \delta \sim O(10)$ is usually sufficient, although the larger this ratio, the better the optimal circuits we would find.

Without loss of generality, and for illustrative purposes, we identify the discrete state corresponding to the $\vert 1 \rangle$ state (hereafter referred to as the ``discrete $\vert 1 \rangle$ state") as the target state of our MDP. To prepare the $|1\rangle$ state starting from any pure 1-qubit state using the gates $RZ$ and $RY$, it is well-known that we require at most a single $RZ$ rotation followed by a single $RY$ rotation. For states lying along the great circle through the $x$ and $z$ axes, we need only a single $RY$ rotation. As a test of this discretized procedure, we investigate whether solving this MDP would be able to reproduce such optimally short gate sequences. We also consider the gateset $\lbrace I, H, T\rbrace$ below, where we include the identity gate to allow for the goal state to ``do nothing" and remain in its state. For simplicity and illustrative purposes, we also include the $S=T^2$ gate in the case of single-qubit state preparation.

\subsection{Reward Structure and Environment Dynamics}
An obvious guess for a reward would be the fidelity $|\langle \phi | \psi\rangle|^2$ between the target state $|\psi\rangle$ and the prepared state $|\phi\rangle$. However, here we consider an even simpler reward structure of assigning $+1$ to the target state, and $0$ to all other states. This allows us to directly relate the length of optimal programs to the value function corresponding to the optimal policy, as we show below. 

To finish our specification of the MDP, we also estimate the environment dynamics $p(s^{\prime}, r | s, a)$. Since our reward structure specifies a unique reward $r$ to every state $s^{\prime} \in \mathcal{S}$, these conditional probabilities reduce to simply $p(s^{\prime} | s,a)$. The discretization of the Bloch sphere implies that the action of a quantum gate $a$ on a discrete state $s=(n,m)$ maps this state to other states $s'=(n',m')$ according to a transition probability distribution $p(s^{\prime} \vert s, a)$. This non-determinism of the effect of the actions occurrs because the discrete states are themselves composed of entire families of continuous quantum states, which are themselves mapped deterministically to other continuous quantum states. However, continuous states from the state discrete state region can land in different discrete final state regions. A simple way to estimate these probabilities is to uniformly sample points on the 2-sphere, determine which discrete state they land in, then perform each of the actions to determine the state resulting from this action. We sample uniformly across the Bloch sphere by sampling $u, v \sim \mathcal{U}[0,1]$, then setting $\theta = cos^{-1}\left(2u -1 \right)$ and $\phi = 2\pi v$. Although other means of estimating these probabilities exist, we find that this simple method works well in practice for the particular problem of single-qubit state preparation.

\begin{figure}[t]
\includegraphics[width=\columnwidth]{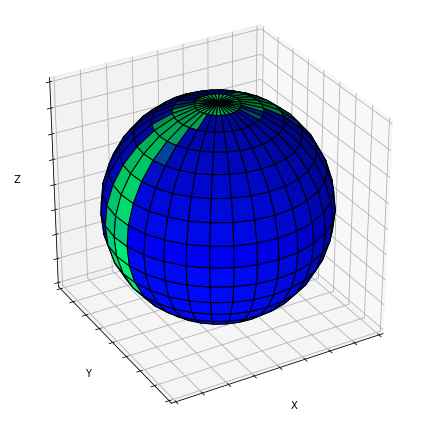}
\caption{\label{fig:rzry} Optimal values for various states on the Bloch sphere using the discrete $RZ$ and $RY$ gates, with a discount factor $\gamma = 0.8$. 
The color of a state corresponds to its optimal value function $V_{\pi^{*}}$, where lighter colors indicate a larger value.
Those colored in green are also exactly the states whose optimal circuits to prepare the discrete $\vert 1 \rangle$ state consist of a single $RY$ rotation, while those in blue are also exactly the ones whose optimal circuits consist of an $RZ$ rotation followed by an $RY$ rotation.}
\end{figure}

\begin{figure}[t]
\includegraphics[width=\columnwidth]{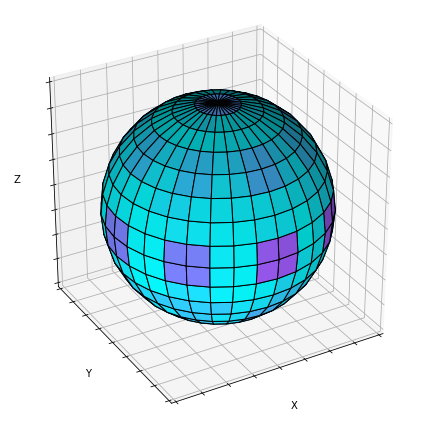}
\caption{\label{fig:ihst} Optimal value landscape across the Bloch sphere using the set of gates $\{I, H, S,T\}$, with a discount factor $\gamma = 0.95$.
The color of a state corresponds to its optimal value function $V_{\pi^{*}}$, where darker colors indicate a larger value.
States distributed around the equator of the Bloch sphere are especially advantageous to start from in order to reach the target $\vert 1 \rangle$ state, as their optimal circuits consist of short sequences of $S$ and $H$ gates.
}
\end{figure}

Note that for the target state $\vert 1 \rangle$, the optimal policy is to just apply the identity, i.e. $RZ(0)$ or $RY(0)$. This action will continue to keep this state in the target state, while yielding $+1$ reward at every time step. This yields an infinite series $V^{\star}(t) = \sum_{k=0}^{\infty} \gamma^{k}$, where $V^{\star} (s) := V_{\pi^{*}}$ and $t$ is the target state, which we can trivially sum to obtain $(1 - \gamma)^{-1}$. This is the highest value of any state on the discretized Bloch sphere. For $\gamma = 0.8$, we obtain $V^{\star} (t) = 5.0$. For some generic state $s \in \mathcal{S}$, we can show that with our reward structure, the optimal value function is given by
\begin{eqnarray}
V^{\star}(s) &=& \sum_{k=0}^{\infty} \gamma^{k} \left( P^{k+1}\right)_{t,s}
\label{eqn:opt-val-fcn}
\end{eqnarray}
where the elements of the matrix $P$ are given by $P_{s',s} = p(s'\vert s,\pi^{\star}(s))$. From Eq. \ref{eqn:opt-val-fcn}, it immediately follows that $V^{\star}(s) \leq V^{\star}(t)$ for all $s \in \mathcal{S}$. The Markov chain produced by the optimal policy has an absorbing state given by the target state, and for some large enough number of steps, all (discrete) states land in this absorbing state. Indeed, the smallest $K$ for which the Markovian process converges to a steady state, such that
\begin{equation}
\left( P^{K}\right)_{s',s} = \delta_{s',t}
\label{eqn:MDP-steady-state}
\end{equation}
for all $s,s' \in \mathcal{S}$ provides an upper bound for the length of the gate sequence that leads from any one discrete state $s$ to the target discrete state $t$. Thus, for the target state itself, $K=0$. Since $\left( P^{k} \right)_{s',s} \leq 1$, for states that are one gate removed from the target state $s_{1}$, we have $V^{\star}(s_1) \leq V^{\star}(t)$, and more generally $V^{\star}(s_{k+1}) \leq V^{\star}(s_k)$. This intimately relates the length of the optimal program to the optimal value function.

The optimal value landscape for the two gatesets are shown in Figs.~(\ref{fig:rzry}) and~(\ref{fig:ihst}). Note that while in the case of the discretized $\lbrace RZ, RY \rbrace$ gates we have a distinguished ring of states along the equator around the $x$-axis that are only a single gate application away from the target state, we have no such continuous patch on the Bloch sphere for the $\lbrace I, H, S, T\rbrace$ gateset, even though there may be indidividual (continuous) states that are only a single gate application away from the target state, e.g. $H\vert 1 \rangle$ for the target state $\vert 1 \rangle$. This shows that states which are nearby on the Bloch sphere need not share similar optimal paths to the target state, given such a gateset.

\subsection{Optimal State Preparation Sequences}
Using policy iteration allows for finding the optimal policy in an MDP. The optimal policy dictates the best action to perform in a given state. We can chain the actions drawn from the optimal policy together to find an optimal sequence of actions, or gates, to reach the target state. In our case, the actions are composed of unitary operations, which deterministically evolve a quantum state (note that we consider noise below in which case the unitary gates are replaced by non-unitary quantum channels). However, due to the discretization, this is no longer true in our MDP, where the states evolve according to the non-trivial probabilities $p(s^{\prime} | s, a)$. The optimal policy is learned with respect to these stochastic dynamics, and not with respect to the underlying deterministic dynamics. In other words, we are imposing a Markovian structure on essentially non-Markovian dynamics. Therefore, if we simply start with some specific quantum state, and apply a sequence of actions drawn from the optimal policy of the discrete states that the evolving quantum states belong to, we might not necessarily find ourselves in the target (discrete) state. For instance, the optimal policy in any one discrete state may be to apply the Hadamard gate, and for a subset of quantum states within that discrete state, this may lead to another discrete state for which the optimal policy is again the Hadamard gate. In such a case, the evolution would be stuck in a loop.

To circumvent this issue, in principle one may allow ``shuffling" of the quantum states within a particular discrete state before evolving them under the optimal policy. However, this may increase the length of the gate sequence and moreover lead to poorer bounds on the fidelity, since $\langle\psi^{\prime}_{f} | \psi_{f}\rangle = \langle\psi^{\prime}_{i} | U^{\dagger}_{1} U^{(1)}_{s} ... U^{(n-1)}_{s} U^{\dagger}_{n} \cdot U_{n} U^{(n)}_{s}  ... U^{(2n-2)}_{s} U_1|\psi_{i}\rangle \neq \langle\psi^{\prime}_{i} | \psi_{i}\rangle$, in general, where the ``shuffling" transformations are given by $U_{s}^{(i)} : \vert \psi \rangle \rightarrow \vert \tilde{\psi}\rangle$ such that $\vert \psi \rangle \sim \vert \tilde{\psi}\rangle$ belong to the same discrete state, while the $U_{i}$ specify (unitary) actions sampled from the optimal policy. On the other hand, without such ``shuffling", the fidelities in the target states from sequences that only differ in their starting states is the same as the fidelities of the starting states, i.e. $\langle\psi^{\prime}_{i} | \psi_{i}\rangle = \langle\psi^{\prime}_{i} | U^{\dagger} U | \psi_{i}\rangle = \langle\psi^{\prime}_{f} | \psi_{f}\rangle$, where $|\psi_{i}\rangle$ and $|\psi^{\prime}_{i}\rangle$ are two different initial pure states that belong to the same initial discrete state, and $U = \prod_i U_i$ is the product of the optimal policies $U_i$.

To avoid such shuffling while still producing convergent paths, we sample several paths that lead from the starting state and terminate in the target (discrete) state, discarding sequences that are larger than some acceptable value, e.g. the length $K$ defined by Eq.~\ref{eqn:MDP-steady-state}, and report the one with the smallest length as the optimal (i.e. shortest) program. Schematically, this can be described in pseudo-code as in Algorithm~\ref{alg:opt-prog1}. This algorithm can be used to generate optimal programs for any given (approximately universal) single-qubit gateset. In our experiments, we found $M$ to be $2$ for the discrete $\{RZ, RY\}$ gateset, and $88$ for the $\{I, H, S, T\}$ gateset, and took $K$ to be 100.

\begin{algorithm}[H]
  \caption{Optimal State Preparation Sequence}
\label{alg:opt-prog1}
\begin{algorithmic}[1]
\State Inputs:-
\State \textsc{Optimal-Policy, Target-State}
\State State space $\mathcal{S}$
\State $K$: length of largest acceptable program
\State $M$: number of sequences to sample
\State
\State Output:-
\State \textsc{Optimal-Programs} for each state
\State
\State Initialize empty array \textsc{Optimal-Programs}
\For{$s\in\mathcal{S}$}
	\State Initialize empty list \textsc{Convergent-Programs}
	\State \textbf{for} $i=1$ to $M$ do
		\State\qquad Converged $\leftarrow False$
		\State\qquad \textbf{while} \textbf{not} Converged
			\State\qquad\qquad State $\leftarrow s$
			\State\qquad\qquad Prog $\leftarrow $ Empty Program list
			\State\qquad\qquad Counter $\leftarrow 0$
			\State\qquad\qquad\textbf{while} Counter $\leq $ $K$
			\State\qquad\qquad\qquad Action $\leftarrow$ \textsc{Optimal-Policy}[State]
			\State\qquad\qquad\qquad Prog.append(Action)
			\State\qquad\qquad\qquad Next-State $\leftarrow$ Env.Step(Action)
			\State\qquad\qquad\qquad State $\leftarrow$ Next-State
			\State\qquad\qquad\qquad Counter $\leftarrow$ Counter + 1
			\State\qquad\qquad\qquad\textbf{if} \textsc{State $=$ Target-State}
			\State\qquad\qquad\qquad\qquad Converged $\leftarrow$ True
			\State\qquad\qquad\qquad\qquad \textsc{Convergent-Programs}.append(Prog)
	\State \textbf{end for}
	\State \textsc{Optimal-Prog} $\leftarrow$ Program with Min length in \textsc{Convergent-Programs}
	\State \textsc{Optimal-Programs}[s] $\leftarrow$ \textsc{Optimal-Prog}
\EndFor
\end{algorithmic}
\end{algorithm}

\subsubsection{Discrete $RZ$ and $RY$ gateset}
In the case of discrete $RZ$ and $RY$ gates, we find what we would expect at most a single $RZ$ rotation followed by a single $RY$ rotation to get from anywhere on the Bloch sphere to the (discrete) $\vert 1 \rangle$ state. For (discrete) states lying along the equatorial ring around the $Y$-axis, we need only apply a single $RY$ rotation. Empirically, we choose a state resolution of $\epsilon = \pi/16$ so that we would find sequences generating the pure $|1\rangle$ state from various discrete states across the Bloch sphere with $\cos^{2} \left( \frac{\pi}{32}\right) \sim 99\%$ fidelity. The optimal programs we find via the preceding procedure for this gateset are composed of programs with lengths either 1 or 2.

\subsubsection{Discrete $\{I, H, T\}$ gateset}
We can also use the procedure described above to obtain approximations to the states $\left( HT\right)^n \vert 0 \rangle$ for integers $n \geq 1$. The unitary $HT$ can be thought of as a rotation by an angle $\theta = 2 \arccos \left( \frac{\cos(7\pi/8)}{\sqrt{2}} \right)$ about an axis $\bm{n} = (n_x, n_y, n_z) = \sqrt{\frac{1}{17}} \left( 5 - 2 \sqrt{2}, 7 + 4 \sqrt{2}, 5 - 2 \sqrt{2} \right)$. The angle $\theta$ has the continued fraction representation
\begin{equation}
\theta = \pi + \frac{\cos\left( \frac{\pi}{8}\right) \sqrt{2 - \cos^2 \left( \frac{\pi}{8}\right)}}{1 + K_{k=1}^{\infty} \frac{-\cos^2 \left( \frac{\pi}{8}\right) \lfloor \frac{1 + k}{2} \rfloor \left( -1 + 2 \lfloor \frac{1+k}{2} \rfloor\right)}{1+ 2 k}}
\end{equation}
which is infinite, and thus the angle $\theta$ is irrational. In the above, we have used the Gaussian notation for the continued fraction
\begin{equation}
    K_{n=1}^{\infty} \frac{a_n}{b_n} = \frac{a_1}{b_1 + \frac{a2}{b2 + \frac{a_3}{b_3 + \dots}}}
\end{equation}
and $\lfloor x \rfloor$ is the flooring operation $x\rightarrow n$ where $n \in \mathbb{Z}$ is the closest integer where $n \leq x$.
The states $(HT)^n \vert 0 \rangle$ lie along an equatorial ring about the axis $\vec{n}$, and no two states $(HT)^n \vert 0 \rangle$ and $(HT)^m \vert 0 \rangle$ are equal for $n \neq m$. Increasing the value of $n$ corresponds to preparing states from among a finite set of states that span the equatorial ring about the axis of rotation. We choose to investigate state preparation up to $n = 10^{10}$. Although as their form makes explicit, these states can be reproduced exactly using $n$ many $H$ and $T$ gates, using our procedure, they can only be obtained up to some fidelity controlled by the discretization as described above. The advantage is that we can obtain good approximations to these states with much fewer gates than $n$. This is illustrated in Table~\ref{tab:htn_states} where short gate sequences can reproduce states of the form $(HT)^n \vert 0 \rangle$ for very large values of $n$ using only a few (between $3$ and $17$ gates). 

\begin{table}[h!]
\begin{center}
\caption{Gate sequences obtained from the optimal policy to approximately produce target states $\ket{\psi_{\text{target}}} = (HT)^n \vert 0 \rangle$. The optimal policy and fidelity are calculated for the noiseless case. The fidelity is defined as $\mathcal{F} = \braket{\psi_{\text{target}}|\psi}$, where $\ket{\psi}$ is obtained from application of the shown gate sequences to the state $\ket{0}$. The sequences are to be read from right to left.}
\label{tab:htn_states}
\begin{tabular}{|c|c|c|}
\hline
$n$ & \text{Gate sequence} & $\mathcal{F}$\\
\hline
$10^2$ & TTHTHTHTH & 0.987 \\
\hline
$10^3$ & TTTHTHTTTH & 0.998\\
\hline
$10^4$ & HTH & 0.992\\
\hline
$10^4$ & HTH & 0.994\\
\hline
$10^6$ & THTTH & 0.998\\
\hline
$10^7$ & HTTTTTHTHTHTHTTTH & 0.998 \\
\hline
$10^8$ & I & 0.999\\
\hline
$10^9$ & I & 0.996\\
\hline
$10^{10}$ & HTTTHTHTHTH & 0.992\\
\hline
\end{tabular}
\end{center}
\end{table}

\section{Noisy state preparation}
Reinforcement learning has previously shown success when applied in the presence of noise ~\cite{An-et-al, Niu-et-al}. Indeed, the ability to learn the effects of a noise channel has apparent practical use when applied to the current generation of noisy quantum computers. These devices are often plagued by errors that severely limit the depth of quantum circuits that can be executed. As full error correction procedures are too resource intensive to be implemented on current hardware, error mitigation methods have been developed to decrease the effect of noise. Methods such as zero-noise extrapolation (ZNE), Clifford data regression (CDR), and probabilistic error cancellation (PEC) involve \emph{post-processing} of circuit results (with and without making use of knowledge about the underlying type of noise)~\cite{temmeErrorMitigationShortDepth2017,2017errormin, mariExtendingQuantumProbabilistic2021,loweUnifiedApproachDatadriven2021}. However, there are also \emph{pre-processing} error mitigation schemes that aim to modify the input circuit in order to reduce the impact of noise. Examples are quantum optimal control methods and dynamical decoupling~\cite{Viola_Knill-PRL-2005,Khodjasteh_Viola-PRL-2009, Abdelhafez_Koch-PRA-2019}. Such techniques attempt to prepare a desired quantum state on a noisy device using circuits (or sequence of pulses) that are different from the ones that would be optimal in the absence of noise. This idea is immediately applicable in our MDP framework as we now demonstrate.

\subsection{State and Action Spaces}
In the presence of noise, the quantum state becomes mixed and is described by a density matrix, which for a single qubit can generally be written as
\begin{equation}
    \rho = \frac{1}{2}(I + r_x X + r_y Y + r_z Z) = \frac{I + \bm{r} \cdot \bm{\sigma}}{2}\,.
\end{equation}
Here, $\bm{r} = (r_x, r_y, r_z)$ are real coefficients called the Bloch vector and $\bm{\sigma} = (X, Y, Z)$ are the Pauli matrices. Since density matrices are semi-definite, it holds that $|\bm{r}| \leq 1$. If $\rho$ is a pure state, then $|\bm{r}| = 1$, otherwise $|\bm{r}| < 1$. Pure states can thus be interpreted as points on the surface of the Bloch sphere, whereas mixed states correspond to points within the Bloch sphere. The maximally mixed state $\rho = I/2$ corresponds to the origin. To find $\bm{r}$, one can calculate the expectation values of each Pauli operator
\begin{align}
    \bm{r} &= \Bigl(\text{Tr}(\rho X), \text{Tr}(\rho Y), \text{Tr}(\rho Z) \Bigr) \\
    &= (r \sin \theta \cos \phi, r \sin \theta \sin \phi, r \cos \theta) \,,
\end{align}
where $r \equiv |\bm{r}| \in [0, 1]$, $\theta \in [0, \pi]$, and $\phi \in [0, 2\pi)$. 

We perform the state discretization analogously to the previous section, but now need to discretize states within the full Bloch ball. To this end, we fix $\epsilon = \pi / k$ and $\delta = 1 / k$ for some positive integer $k$. Now the set of points $n\epsilon \le \theta \le (n+1)\epsilon$, $m\epsilon \le \phi \le (m+1)\epsilon$, and $l \delta \le r \le (l + 1)\delta$ for integers $1 \leq n\leq k-2$, $0 \leq m \leq 2k-1$, and $0 \leq l \le k-1$ constitute the same discrete state $s = (n, m, l)$ in the MDP. As before, the polar regions $n = 0, k-1$ are special as these regions are independent of $\phi$, i.e. they are described by the set of integers $s = (n, m=0, l)$. 
This discretization corresponds to nesting concentric spheres and setting the discrete MDP states $s$ to be the 3-dimensional regions between them. 

Let us now introduce the action space $\mathcal{A}$ in the presence of noise. We model noisy gates using a composition of a unitary gate $U$ and a noisy quantum channel described by a set of Kraus operators
\begin{equation}
    \mathcal{E}(\rho) = \sum_k E_k \rho E_k^\dag
\end{equation}
with $\sum_k E_k E_k^\dag = \mathbb{I}$. Application of a noisy quantum channel can shrink the magnitude $r$ of the Bloch vector as the state becomes more mixed. Evolution under a unitary gate $U$ in this noisy channel results in
\begin{equation}
    U\mathcal{E}(\rho) = \sum_k U E_k \rho E_k^\dag U^\dag 
\end{equation}
We here again consider the discrete gateset $U \in \{I,H,T\}$. Once we specify the type of noise via a set of Kraus operators, its sole effect on our description of the MDP is to change the transition probability distributions $p(s^{\prime} \vert s, a)$. While noise can change the optimal policies, we may nevertheless solve for the optimal policies using the exact same procedure that we used in the noiseless case. In the following, we compare the resulting shortest gate sequences found by an agent that was trained using the noisy transition probabilities $p$ and compare them to those found by an agent lacking knowledge of the noise channel. 

The noise observed in current quantum computers is to a good approximation described by amplitude damping and dephasing channels. The amplitude damping channel is described by the two Kraus operators
\begin{equation}
    E_0 = \left( \begin{array}{ccc}
    0 & \sqrt{\gamma} \\
    0 & 0 \end{array} \right) \,,\,  
    E_1 = \left( \begin{array}{ccc}
    1 & 0 \\
    0 & \sqrt{1-\gamma} \end{array} \right) \ 
\end{equation}
with $\ 0 \le \gamma \le 1$. Physically, we can interpret this channel as causing a qubit in the $\ket{1}$ state to decay to the $\ket{0}$ state with probability $\gamma$. In current quantum computing devices, the relaxation time, $T_1$, describes the timescale of such decay processes. For a given $T_1$ time and a characteristic gate execution time $\tau_g$, we parametrize
\begin{equation}
\label{eq:gamma_T_1}
    \gamma = 1 - e^{-\tau_g/T_1}\,.
\end{equation}
Note that application of the amplitude damping channel also leads to dephasing of the off-diagonal elements of the density matrix in the $Z$ basis with timescales $T_2 = 2 T_1$. 

The dephasing channel takes the form
\begin{equation}
    \mathcal{E}(\rho) = (1-p)\rho + pZ\rho Z^\dagger
\end{equation}
and is described by the Kraus operators $A_0 = \sqrt{1-p} \, \openone$ and $A_1 = \sqrt{p} Z$. This channel leads to pure phase damping of the off-diagonal terms of the density matrix in the $Z$ basis. It is described by a dephasing time, $T_2$. 

We use Pyquil~\cite{smith2016practical} (specifically the function \texttt{damping\_after\_dephasing} from \texttt{pyquil.noise}) to construct a noise channels consisting of a composition of these two noise maps by specifying the $T_1$ and $T_2$ times as well as the gate duration $\tau_g$. The amplitude damping parameter $\gamma$ is set by $T_1$ using Eq.~\eqref{eq:gamma_T_1}. Since this also results in phase damping of the off-diagonal terms of the density matrix (in the $Z$ basis), the dephasing time $T_2$ is upper limited by $2 T_1$. We thus parametrize the dephasing channel parameter (describing any additional pure dephasing) as  
\begin{equation}
    \label{eq:p_T_2}
    p = \frac12 \bigl( 1- e^{-\tau_g[1/T_2 - 1/(2T_1)] } \bigr) \,.
\end{equation}
The dephasing channel thus describes dephasing leading to $T_2 < 2 T_1$ and acts trivially if $T_2$ is at its upper bound $T_2 = 2 T_1$. In the following, we consider $T_1 = T_2$ such that the dephasing channel acts non-trivially on the quantum state. 

\subsection{Reward Structure and Environment}
For the reward structure of our noisy state preparation, we consider the purity of the state when calculating the reward. This is to account for the fact that there may be no gate sequence that results in a state with a high enough purity to land in the pure goal state. As such, assigning a reward of $+1$ to the pure target state and $0$ to all other states can lead to poor convergence. 
We assign the reward as follows: the pure target state $\rho_\text{target}$ is one of the MDP states $(n_{\text{target}}, m_{\text{target}}, k-1)$, where $r = 1$ and thus $l = k-1$. Considering a state $\rho$ in the MDP state $(n',m',l')$. If $n' = n_{\text{target}}$ and $m' = m_{\text{target}}$, we assign a reward of $l'/k$. Otherwise, we assign a reward of 0. With this construction, we reward reaching a state in the target direction (i.e. with correct angles $\theta, \phi$) using a reward amount that is proportional to the purity of the state. We thus also reward gate sequences that do not end up in the pure goal state, while still encouraging states with higher purity. 

One can expect the optimal value function for a noisy evolution to take on smaller values due to the fact that the rewards are smaller. Indeed, it can be easily verified that for a simplified noise model consisting of only a depolarizing quantum channel $\mathcal{E}(\rho) = (1 - p)\rho + \frac{p}{3}(X\rho X + Y\rho Y + Z\rho Z)$, the resulting optimal value function is simply uniformly shrunk compared to the optimal value function of a noiseless MDP. Since the change is uniform across all values, the optimal policy is unchanged from the noiseless setting. This is no longer the case for realistic noise models such as described by amplitude and dephasing quantum channels, in which case we rederive the optimal policy using policy iteration. 

This requires updating the conditional probability distributions $p(s'|s,a)$ by performing Monte-Carlo simulations as before by drawing random continuous states from within a given discrete states, applying deterministic noisy gates, and recording the obtained discrete states. Now we find transitions to states $s'$ with lower purity than the initial state $s$. Note that the randomness of the probability distribution $p$ arises solely from the randomly sampled continuous quantum states to which we apply the noisy gate actions. The randomness due to noise is fully captured within the mixed state density matrix description of quantum states. 

\subsection{Optimal Noisy State Preparation Sequences}
We now consider the task of approximating states of the form $(HT)^n\ket{0}$ for $n \gg 1$, starting from the  state $\ket{0}$, using a gate set $\{I, H, T\}$ in the presence of noise. We use the MDP formulation with transition probabilities $p(s'|s,a)$ obtained in the presence of amplitude and dephasing noise channels. We find the optimal policy using policy iteration that yields optimal gate sequences via Algorithm~(\ref{alg:opt-prog1}). 
\begin{table*}[t]
    \centering
    \begin{tabular}{|c || c | c | c | c |}
         \hline
         $n$ & \text{Noiseless MDP} & $\mathcal{F}$ & \text{Noisy MDP} & $\mathcal{F}$ \\ [0.4ex]
         \hline \hline
         $10^2$ & TTHTHTHTH & 0.774 & HIIHTH & 0.882 \\ [0.1ex] 
         \hline
         $10^3$ & TTTHTHTTTH & 0.652 & HTHTTTTHH & 0.820 \\ [0.1ex] 
         \hline
         $10^4$ & HTH & 0.843 & HTHTTTTHH & 0.820\\ [0.1ex]
         \hline
         $10^5$ & HTH & 0.824 & TTTTTTH & 0.869 \\ [0.1ex]
         \hline
         $10^6$ & THTTH & 0.735 & HTTHTIHTH & 0.863 \\ [0.1ex] 
         \hline
         $10^7$ & HTTTTTHTHTHTHTTTH & 0.600 & HTHTTTTTTH & 0.800 \\ [0.1ex]
         \hline
         $10^8$ & I & 0.999 & I & 0.999 \\ [0.1ex]
         \hline
         $10^9$ & I & 0.999 & I & 0.999 \\ [0.1ex]
         \hline
         $10^{10}$ & HTTTHTHTHTH & 0.702 & HTHITTTTTTH & 0.806 \\ [0.1ex]
         \hline
    \end{tabular}
    \caption{Shortest gate sequences and noisy fidelities $\mathcal{F}$ produced by the optimal policies $\pi^*$ of noiseless MDP (columns 2 and 3) and noisy MDP (columns 4 and 5). The gate sequences should be read right to left. The noise is characterized by $T_1$ = $T_2$ = $1 \,\mu$s and the gate time is set to $\tau_g = 200$~ns. While this corresponds to a noise level that is stronger than in current day NISQ hardware, where $T_1, T_2 \approx 100 \mu$s, these parameters yield sufficiently strong noise to highlight differences in the optimal gate sequences. The fidelities (with the target state) are obtained by preparing mixed states using the shown gate sequences applied to $\ket{0}$ in the presence of noise. Note that the states generated by $(HT)^n\ket{0}$ for $n = 3,4$ have an overlap fidelity of $0.99$. This is also the case for $n = 7, 10$. This  explains the similarity of gate sequences found for these cases.}
    \label{table:optimalgateseq}
\end{table*}

\begin{figure}[t]
    \centering
    \includegraphics[width=\columnwidth]{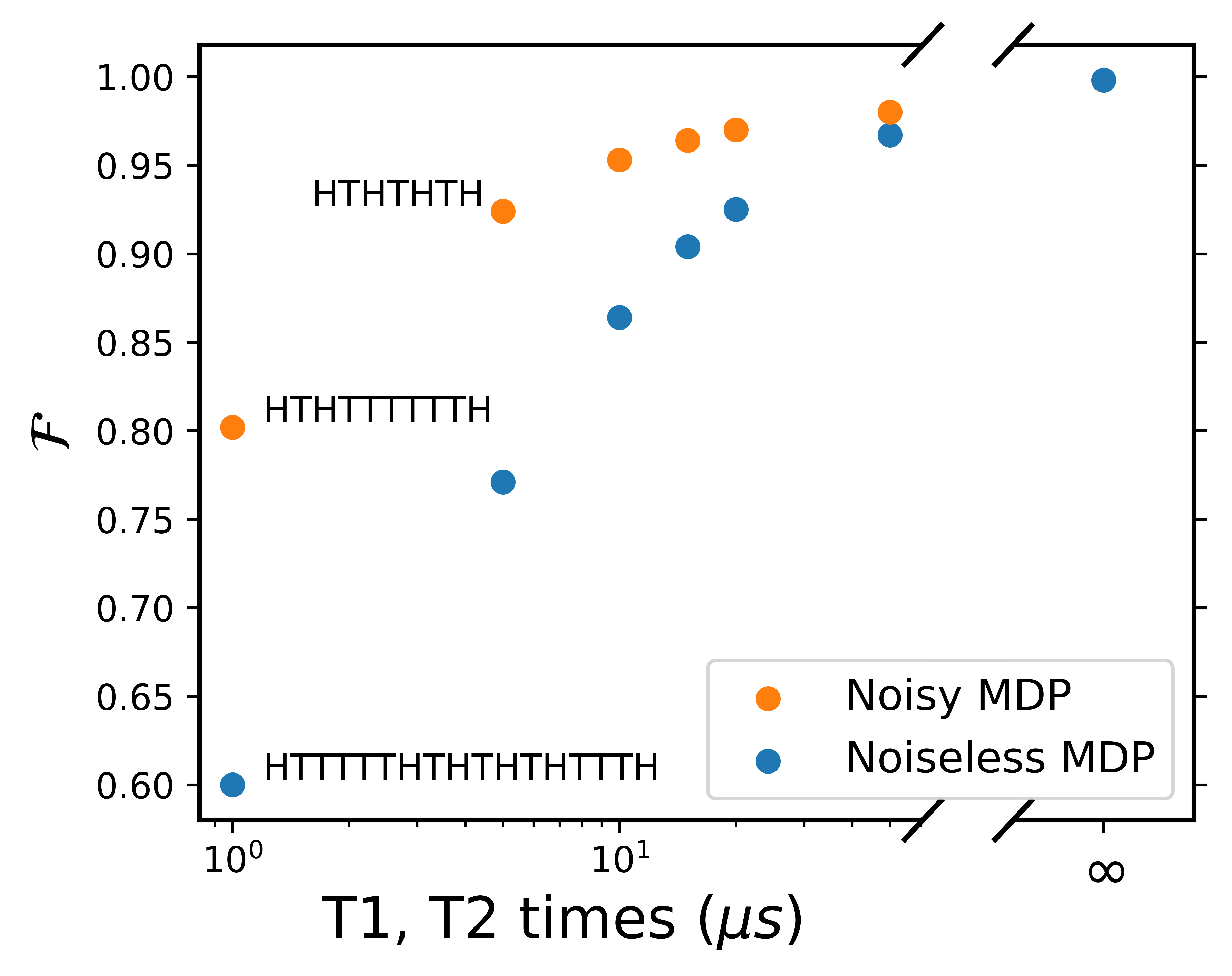}
    \caption{Fidelity $\mathcal{F}$ of the state $\sigma$ prepared using optimal gate sequences with target state $\rho_{\text{target}} = (HT)^n\ket{0}$ for fixed $n=10^7$ as a function of noise strength $T_1 = T_2$. The shortest gate sequences (indicated in the figure) are produced by optimal policies $\pi^*_{\text{noisy}}$ (orange) and $\pi^*_{\text{noiseless}}$ (blue) of noisy and noiseless MDPs, respectively. The noisy policy gives gate sequences that are different from the noiseless case, which consistently yield higher fidelities. The optimal noisy gate sequence is $HTHTHTH$ for all times $T_1=T_2 \geq 60 \mu$s. We fix the gate time to $\tau_g =200$~ns when generating the Kraus operators as defined by Eqs.~\eqref{eq:gamma_T_1} and~\eqref{eq:p_T_2}. For each value of $T_1, T_2$, we generate the transition probabilities $p(s'|s,a)$ according to the corresponding noise map and use policy iteration to find the optimal policy. The fidelity is then calculated by applying the gate sequence found by both the noisy and noiseless MDPs to $\ket{0}$ in the error channel for that specific value of $T_1$, $T_2$. The point at infinity represents the noiseless case, and corresponds to the transition probabilities learned by the noiseless MDP. }
    \label{fig:t1t2vsfidelity}
\end{figure}

In Table~\ref{table:optimalgateseq}, we present results up to $n=10^{10}$ that includes the shortest gate sequences found by the optimal policies of noisy and noiseless MDP. We also compare the final state fidelities produced by these optimal circuits.  The fidelities $\mathcal{F}$ that are listed in the table are found by applying the optimal gate sequences for a given $n$ to the exact state $\ket{0}$. Since the resulting states are mixed, we calculate the fidelity between the target state $\sigma$ and the state resulting from an optimal gate sequence $\rho$ as 
\begin{equation}
   \mathcal{F}(\rho, \sigma) = \big( \text{tr} \sqrt{\sqrt{\rho} \sigma \sqrt{\rho}} \big)^2\,. 
\end{equation}
We list both the gate sequences found by the noiseless MDP, the agent whose underlying probability distribution $p(s'|s,a)$ is constructed from exact unitary gates, and the noisy MDP, whose transition probabilities are generated from noisy gates considering combined amplitude damping and dephasing error channels. We set the relaxation and dephasing times to $T_1 = T_2 = 1 \mu$s and the gate time to  $\tau_g = 200$~ns. While the value for $\tau_g$ is typical for present day superconducting NISQ hardware, the values of $T_1, T_2$ are about two orders of magnitude shorter than typical values on today's NISQ hardware, where $T_1, T_2 \simeq 100 \mu$s. We choose such stronger noise values in order to highlight the difference in gate sequences (and resulting fidelities) produced by the optimal policies $\pi^*$ for noisy and noiseless MDPs. We expect that this result is generic and robust when considering MDPs for multiple qubits, where two-qubit gate errors are expected to lead to more pronounced noise effects.

The results in Table~\ref{table:optimalgateseq} demonstrate that even in the presence of (strong) noise, the noisy MDP is able to provide short gate sequences that approximate the target state reasonably well. Importantly, for all values of $n$ shown (except for $n=10^4$), the optimal policy of the noisy MDP $\pi^*_{\text{noisy}}$ yields a gate sequence that results in a higher fidelity than the gate sequence obtained from $\pi^*_{\text{noiseless}}$ of the noiseless MDP (if applied in the presence of noise). This shows that noise can be mitigated by adapting the gate sequence according to the noise experienced by the qubit. Solving for the optimal policies of a noisy MDPs are a convenient approach to finding such adapted quantum circuits.  

In Fig.~\ref{fig:t1t2vsfidelity} we compare the gate sequences and fidelities obtained from the optimal policies of noisy and noiseless MDPs for a fixed value of $n=10^7$ as a function of $T_1 = T_2$. We observe the noisy MDP to outperform the noiseless MDP for all noise strengths. This indicates that by learning the noise channel, the agent can adapt to the noise and find gate sequences that yield higher fidelities in that channel. Note that if we applied the gate sequences found by the noisy MDP in a noiseless setting, they would yield lower fidelities than gate sequences produced by the optimal policy of a noiseless MDP.
Based on these result, we conclude that dynamic programming and reinforcement learning methods provide a powerful and generic way to perform pre-processing error mitigation by identifying optimal gate sequences for qubit state preparation in the presence of noise. Future work should be directed towards exploring these approaches for two and more coupled qubits.

\section{Compilation of single-qubit gates}
In the previous sections, we considered an agent-environment interaction in which we identified Hilbert space as the state space, and the space of $SU(2)$ gates as the action space. Shifting our attention to the problem of quantum gate compilation, we now identify both the state and action spaces with the space of $SU(2)$ matrices, where for convenience we ignore an overall $U(1)$ phase from the true group of single-qubit gates $U(2)$. We first consider an appropriate coordinate system to use, and discuss why the quaternions are better suited to this task than Euler angles. We focus exclusively on the gateset $\lbrace I, H, T \rbrace$, and modify the reward structure slightly so that we now have to work with the probabilities $p(s^{\prime}, r | s , a)$ instead of the simpler $p(s^{\prime} | s, a)$ as in the previous section. We present empirical results for a few randomly chosen (special) unitaries.

\subsection{Coordinate system}
We consider the gateset $\{I, H, T\}$. We include the identity in our gate set since we would like the target state to possess the highest value, and have the agent do nothing in the target state under the optimal policy. Because we would like to remain in the space of $SU(2)$ matrices, we define $H = RY(\pi/2) RZ(\pi)$, which differs from the usual definition by an overall factor of $i$, and $T=RZ(\pi/4)$. Note that owing to our alternative gate definitions, we have that $H^2 = T^8 = -1 \neq 1$ so that we may obtain up to 3 and 15 consecutive applications of $H$ and $T$ respectively in the optimal program. Next, we choose an appropriate coordinate system. One choice is to parametrize an arbitrary $U\in SU(2)$ using the ZYZ-Euler angle decomposition. Under this parametrization, given some $U \in SU(2)$
\begin{equation}
U = U(a, b, c, d) = \begin{pmatrix}
a + ib & c + id \\
-c + id & a - ib
\end{pmatrix}
\label{eqn-U}
\end{equation}
such that $a^2 + b^2 + c^2 + d^2 = 1$, we can write $U = RZ(\alpha) RY(\beta) RZ(\gamma)$ with
\begin{eqnarray}
\alpha &=& \alpha(a, b, c, d) = \arctan{(-b/a)} + \arctan{(-d/c)} \nonumber \\
\beta &=& \beta(a, b, c, d) = 2 \arccos(\sqrt{a^{2} + b^{2}}) \\
\gamma &=& \gamma(a, b, c, d) = \arctan(-b/a) - \arctan(-d/c) \nonumber
\end{eqnarray}
for some angles $\alpha$, $\beta$ and $\gamma$. Note that for $\beta = 0$, we have a continuous degeneracy of choices in $\alpha$ and $\gamma$ to specify some $RZ(\delta)$ with $\alpha + \gamma = \delta$. However, the transformations above will conventionally fix this to $\alpha = \gamma = \delta/2$.

Under the action of $T$, i.e. $T: U \rightarrow U^{\prime} = T\, U = RZ(\alpha^{\prime}) RY(\beta^{\prime}) RZ(\gamma^{\prime})$, or equivalently $T: (\alpha, \beta, \gamma) \rightarrow (\alpha^{\prime}, \beta^{\prime}, \gamma^{\prime})$, the ZYZ-coordinates transform rather simply as $\alpha^{\prime} = \alpha + \pi/4$, $\beta^{\prime} = \beta$, $\gamma^{\prime} = \gamma$. Under a similar action of $H$ however, the coordinates transform non-trivially. The matrix entries, on which these parameters depend, transform as
\begin{eqnarray}
a^{\prime} &=& \frac{1}{\sqrt{2}} \left[ \sin\left(\frac{\alpha - \gamma}{2}\right) \sin\left( \frac{\beta}{2}\right) - \sin\left(\frac{\alpha + \gamma}{2}\right) \cos\left( \frac{\beta}{2}\right)\right] \nonumber \\
b^{\prime} &=& \frac{-1}{\sqrt{2}} \left[ \cos\left(\frac{\alpha - \gamma}{2}\right) \sin\left( \frac{\beta}{2}\right) + \cos\left(\frac{\alpha + \gamma}{2}\right) \cos\left( \frac{\beta}{2}\right)\right] \nonumber \\
c^{\prime} &=& \frac{1}{\sqrt{2}} \left[ \sin\left(\frac{\alpha - \gamma}{2}\right) \sin\left( \frac{\beta}{2}\right) + \sin\left(\frac{\alpha + \gamma}{2}\right) \cos\left( \frac{\beta}{2}\right)\right] \nonumber \\
d^{\prime} &=& \frac{1}{\sqrt{2}} \left[ \cos\left(\frac{\alpha - \gamma}{2}\right) \sin\left( \frac{\beta}{2}\right) - \cos\left(\frac{\alpha + \gamma}{2}\right) \cos\left( \frac{\beta}{2}\right)\right] \nonumber \\
& &
\end{eqnarray}
This is a non-volume preserving operation for which
\begin{eqnarray}
\det(J) &=& \frac{\sin(\beta)}{\sqrt{1 - \cos^2(\alpha) \sin^2(\beta)}}
\end{eqnarray}
where $J$ denotes the Jacobian of the transformation from $(\alpha, \beta, \gamma)$ to $(\alpha^{\prime}, \beta^{\prime}, \gamma^{\prime})$ under the action of $H$, and which diverges for values of $\alpha$ and $\beta$ such that $\cos(\alpha)\sin(\beta) = \pm 1$. This implies that for such pathological values, a unit hypercube in the discretized $(\alpha, \beta, \gamma)$ space gets mapped to a region that covers indefinitely many unit hypercubes in the discretized $(\alpha^{\prime}, \beta^{\prime}, \gamma^{\prime})$ space. In turn, this means that a single state $s$ gets mapped to an unbounded number of possible states $s^{\prime}$, causing $p(s^{\prime}\vert s, a=H)$ to be arbitrary small. This may prevent the agent from recognizing an optimal path to valuable states, since even if the quantity $(r + \gamma V_{\pi}(s^{\prime}))$ is particularly large for some states $s^{\prime}$, this quantity gets multiplied by the negligible factor $p(s^{\prime}\vert s, a=H)$, and therefore has a very small contribution in an update rule such as Eq (\ref{eq:bellman-iterate}).

These problems can be overcome by switching to using quaternions as our coordinate system. Unlike the ZYZ-Euler angles, the space of quaternions is in a one-to-one correspondence with $SU(2)$. Given some $U \in SU(2)$ as in Eq (\ref{eqn-U}), the corresponding quaternion is given simply as $q = (a, b, c, d)$. Under the action of $T$, its components transform as
\begin{eqnarray}
a^{\prime} &=& a \cos\left( \frac{\pi}{8}\right) + b \sin\left( \frac{\pi}{8} \right) \nonumber \\
b^{\prime} &=& b \cos\left( \frac{\pi}{8}\right) - a \sin\left( \frac{\pi}{8} \right) \nonumber \\
c^{\prime} &=& c \cos\left( \frac{\pi}{8}\right) + d \sin\left( \frac{\pi}{8} \right) \nonumber \\
d^{\prime} &=& d \cos\left( \frac{\pi}{8}\right) - c \sin\left( \frac{\pi}{8} \right)
\end{eqnarray}
while under the action of $H$, its components transform as
\begin{eqnarray}
a^{\prime} &=& \frac{b+d}{\sqrt{2}} \nonumber \\
b^{\prime} &=& \frac{c-a}{\sqrt{2}} \nonumber \\
c^{\prime} &=& \frac{d-b}{\sqrt{2}} \nonumber \\
d^{\prime} &=& -\frac{a+c}{\sqrt{2}}
\label{eqn:quat-H}
\end{eqnarray}
and $\det(J_{(T)}) = \det(J_{(H)}) = 1$ for the Jacobians associated with both transformations, so that these operations are volume-preserving on this coordinate system. In turn, this implies that a hypercube with unit volume in the discretized quaternionic space gets mapped to a region with unit volume.

For the purposes of the learning agent, this means that the total number of states that can result from acting with either $T$ or $H$ is bounded above. Suppose we choose our discretization such that the grid spacing along each of the 4 axes of the quaternionic space is the same. Then, since a $d$-dimensional hypercube can intersect with at most $2^d$ equal-volume hypercubes, a state $s$ can be mapped to at most $16$ possible states $s^{\prime}$. While this is certainly better than the pathological case we noted previously using the ZYZ-Euler angles, one could ask if it is possible to do better and design a coordinate system such that a state gets mapped to at most one other state.

One possible approach to make the environment dynamics completely deterministic is to consider a discretization $q = (n_1 \Delta, n_2 \Delta, n_3 \Delta, n_4 \Delta)$ where $n_1, n_2, n_3, n_4 \in \mathbb{Z}$, and choose $\Delta$ such that the transformed quaternion can also be described similarly as $q^{\prime} = (n^{\prime}_1 \Delta, n^{\prime}_2 \Delta, n^{\prime}_3 \Delta, n^{\prime}_4 \Delta)$, and try to ensure that $n^{\prime}_1$, $n_2^{\prime}$, $n_3^{\prime}$, $n_4^{\prime}$ are also integers. Essentially this would mean that corners of hypercubes map to corners of hypercubes, so that discretized states map uniquely to other discretized states. However, consider the transformation under $H$, Eq. (\ref{eqn:quat-H}). For this transformation, requiring $a^{\prime} = (b + d) / \sqrt{2} = (n_2 + n_4) \Delta / \sqrt{2}$ to equal $n^{\prime}_1 \Delta$ in turn requires that $n^{\prime}_1 = k / \sqrt{2}$, for some $k \in \mathbb{Z}$ (and similarly for the other components). This implies that $n^{\prime}_1$ cannot be an integer, and so the map given with this gateset over this discretized coordinate system cannot be made deterministic in this manner. Nevertheless, we find that our construction is sufficient to solve the MDP that we have set up.

\subsection{Reward Structure and Environment Dynamics}
Some natural measures of overlap between two unitaries include the Hilbert-Schmidt inner product $\text{tr}(U^{\dagger}V)$, and since we work with quaternions, the quaternion distance $|q - q^{\prime}|$. However, neither does the Hilbert-Schmidt inner product monotonically increase, nor does the quaternion distance monotonically decrease, along the shortest $\lbrace H, T \rbrace$ gate sequence. As an example, consider the target quaternion $q^{\star} = \left[-0.52514, -0.38217, 0.72416, 0.23187\right]$ from Table (\ref{tab:compilation-seqs}) with shortest compilation sequence $HTTTTHTHHH$ (read right to left) satisfying $|q - q^{\star}| < 0.3$, where $q$ is the prepared quaternion via the sequence. After the first $H$ application, $|q - q^{\star}| \sim 1.34$, which drops after the second $H$ application to $|q - q^{\star}| \sim 0.97$, and then rises again after the third $H$ applciation to $|q - q^{\star}| \sim 1.49$, before eventually falling below the threshold error. Similarly, the Hilbert-Schmidt inner product starts at $\sim 0.21$, rises to $\sim 1.05$, then falls to $\sim -0.21$ before eventually becoming $\sim 1.96$. On the other hand, we showed previously how assigning a reward structure of $+1$ to some target state, and $0$ to all other states, made it possible to relate the optimal value function to the length of the optimal path.

Instead of specifying a reward of $+1$ in some target state and $0$ in every other state however, we now assign a reward of $+1$ whenever the underlying unitary has evolved to within an $\epsilon$-net approximation of the target unitary. Since we work with quaternions, we specify this as obtaining a reward of $+1$ whenever the evolved quaternion $q$ satisfies $|q - q^{\star}| < \epsilon$, for some $\epsilon > 0$ and $q^{\star}$ is the target quaternion, and $0$ otherwise. We note that the Euclidean distance between two vectors $(a, b, c, d)$ and $(a + \Delta_{\text{bin}}, b + \Delta_{\text{bin}}, c + \Delta_{\text{bin}}, d + \Delta_{\text{bin}})$ equals $2\Delta_{\text{bin}}$, however both those vectors cannot represent quaternions, since only either one of them can have unit norm. Nevertheless, this sets a size of discrete states, and we require that $\epsilon$ be comparable to this scale, setting $\epsilon = 2\Delta_{\text{bin}}$ in practice. This requirement comes from the fact that in general, the $\epsilon$-net could cover more than one state, so that we now need to estimate the probabilities $p(s^{\prime}, r | s, a)$, in contrast to the scenario where a state uniquely specifies the reward. Demanding that $\epsilon \sim \Delta_{\text{bin}}$ ensures that $p(s^{\prime}, r=1 | s, a)$ does not become negligibly small.

We could estimate the dynamics by uniformly randomly sampling quaternions, track which discrete state the sampled quaternions belong to, evolve them under the actions and track the resultant discrete state and reward obtained as a result, just as we did in the previous section. However, here we now estimate the environment dynamics by simply rolling out gate sequences. Each rollout is defined as starting from the identity gate, then successively applying either an $H$ or $T$ gate with equal probability until some fixed number $K$ of actions have been performed. The probabilities for the identity action $p(s^{\prime}, r | s, a = I)$ are simply estimated by recording that $(s^{\prime}, a = I)$ led to $(s^{\prime}, r)$ at each step that we sample $(s^{\prime}, r)$ when performing some other action $a \neq I$ in some other state $s \neq s^{\prime}$. The number of actions per rollout $K$ is set by the desired accuracy, which the Solovay-Kitaev theorem informs us is $O(\text{polylog}(1/\epsilon))$ \cite{Solovay-Kitaev-algo}, and in our case has an upper bound given by Eq.~\ref{eqn:MDP-steady-state}. Estimating the environment dynamics in this manner is similar in spirit to off-policy learning in typical reinforcement learning algorithms, such as $Q$-learning \cite{Sutton-barto}.

\subsection{Optimal Gate Compilation Sequences}
Solving the constructed MDP through policy iteration, we arrive at the optimal policy just as before. We now chain the optimal policies together to form optimal gate compilation sequences, accounting for the fact that while the dynamics of our constructed MDP is stochastic, the underlying evolution of the unitary states is deterministic. The procedure we use for starting with the identity gate and terminating, with some accuracy, at the target state is outlined in pseudo-code in Algorithm~\ref{alg:opt-prog}, where the length of the largest sequence $K$ is dictated by Eq.~\ref{eqn:MDP-steady-state}, and in our experiments we took 100 rollouts.
\begin{algorithm}[H]
\caption{Optimal Gate Compilation Sequence}
\label{alg:opt-prog}
\begin{algorithmic}[1]
\State Inputs:-
\State \textsc{Optimal-Policy}
\State Transition probabilities $p(s^{\prime}, r \vert s, a)$
\State \textsc{Reward} structure
\State $K$: Largest acceptable length of gate sequence
\State Number of rollouts
\State
\State Output:-
\State \textsc{Optimal-Sequence}
\State
\State Initialize empty list \textsc{Action-Rollouts}
\For{each rollout}
	\State Initialize empty list \textsc{Action-Sequence}
	\State Initialize \textsc{State} to Identity gate
	\State \textsc{Counter} $\leftarrow$ 0
	\State \textbf{while} \textsc{Counter} $<$ $K$
	\State\qquad \textsc{Action} $\leftarrow$ \textsc{Optimal-Policy[State]}
	\State\qquad \textsc{Action-Sequence}.append(\textsc{Action})
	\State\qquad Sample \textsc{(Next-State, Reward)} from estimated $p(s^{\prime}, r \vert s, a)$
	\State\qquad \textsc{State} $\leftarrow$ \textsc{Next-State}
	\State\qquad \textbf{if} \textsc{Reward} $=$ \textsc{1}
	\State\qquad\qquad \textbf{break}
	\State\textbf{if} \textsc{Action-Sequence} is not empty
	\State\qquad\textsc{Action-Rollouts}.append(\textsc{Action-Sequence})
\EndFor\\
\textsc{Optimal-Sequence} $\leftarrow$ Minimum length \textsc{Action-Sequence} in \textsc{Action-Rollouts} that satisfies precision bound
\end{algorithmic}
\end{algorithm}

The accuracy with which we would obtain the minimum length action sequence in Algorithm (\ref{alg:opt-prog}) need not necessarily satisfy the bound $\epsilon$ set by the reward criterion, $r=1$ for $|q - q^{\star}| < \epsilon$, for reasoning similar to the shuffling discussed in the context of state preparation above. This is why we require Algorithm (\ref{alg:opt-prog}) to report the minimum length action sequence that also satisfies the precision bound. In practice, we found that this was typically an unnecessary requirement and even when the precision bound was not satisfied, the precision did not stray too far from the bound. It should be emphasized that due to the shuffling effect, there is no a priori guarantee that optimal-sequence returned by Algorithm (\ref{alg:opt-prog}) need even exist, since the precision bound is not guaranteed to exist, and the only bound we can safely set is $|q - q^{\star}| \lesssim \Delta_{\text{bin}} k$, where $k$ is the number of actions in the sequence that prepares $q$. In practice however, we find the algorithm to work quite well in producing optimal sequences that correspond to the shortest possible gate sequences to prepare the target quaternions $q^{\star}$.

To benchmark the compilation sequences found by this procedure, we find shortest gate sequences for compilation to some specified precision using a brute-force search that yields the smallest gate sequence that satisfies $|q - q^{\star}| < \epsilon$ for some $\epsilon > 0$ with the smallest value of $|q - q^{\star}|$, where $q$ is the prepared quaternion and $q^{\star}$ is the target quaternion. This brute-force procedure can be described in pseudo-code as in Algorithm~\ref{alg:brute-force}.
\begin{algorithm}[H]
\caption{Shortest Gate Compilation Sequence}
\label{alg:brute-force}
\begin{algorithmic}[1]
\State Inputs:-
\State Target unitary $q^{\star}$
\State \textsc{Found} = \textsc{False}
\State Target accuracy $\epsilon$
\State
\State Output:-
\State \textsc{Shortest-Sequence}
\State
\State \textsc{Found} = \textsc{False}
\State \textbf{while not} \textsc{Found}
\State\qquad Initialize empty list \textsc{Quaternion-Distances}
\State\qquad \textsc{Sequences} $\leftarrow$ $2^n$ sequences of $\lbrace H, T \rbrace$
\State\qquad \textbf{for} \textsc{Seq} in \textsc{Sequences}
\State\qquad\qquad Evolve identity quaternion according to \textsc{Seq}
\State\qquad\qquad \textsc{Quaternion-Distances}.append(\textsc{$|q - q^{\star}|$})
\State\qquad \textbf{if} \textsc{Min(Quaternion-Distances)} $< \epsilon$
\State\qquad\qquad \textsc{Found} $\leftarrow$ \textsc{True}
\State\qquad\qquad \textsc{Shortest-Sequence} $\leftarrow$ \textsc{Seq} with \textsc{Min(Quaternion-Distances)}
\end{algorithmic}
\end{algorithm}

As an experiment, we drew 30 (Haar) random $SU(2)$ matrices, and found their compilation sequences from Algorithms (\ref{alg:opt-prog}) and (\ref{alg:brute-force}). We set $\epsilon = 2 \Delta_{\text{bin}} = 0.3$, estimated the environment dynamics using 1000 rollouts, each rollout being 50 actions long, and each action being a uniform draw between $H$ and $T$. The findings are presented in Table (\ref{tab:compilation-seqs}), where the sequences are to be read right to left. We find that although the two approaches sometimes yield different sequences, the two sequences agree in their length and produce quaternions that fall within $\epsilon$ of the target quaternion. We expect in general that the two approaches will produce comparable length sequences and target fidelities, though not necessarily equal.

\begin{table*}[t]
\begin{center}
\caption{Compilation Gate Sequences from MDP and Brute-Force (BF) using $\epsilon = 0.3$}. Note the definitions of $H = RY(\pi/2) RZ(\pi)$ and $T=RZ(\pi/4)$, as described in the main text.
\label{tab:compilation-seqs}
\begin{tabular}{|c|c|c|c|c|}
\hline
$q^{\star}$ & \text{MDP Gate Sequence} & \text{Brute-Force Gate Sequence} & $|q_{MDP} - q^{\star}|$ & $|q_{BF} - q^{\star}|$\\
\hline
[-0.54981  0.35852  0.41549  0.62972] & THTTH & THTTH & 0.19996 & 0.19996\\
\hline
[-0.76688  0.32823 -0.37129  0.4078] & HTHT & HTHT & 0.2483 & 0.2483\\
\hline
[-0.52514 -0.38217  0.72416  0.23187] & HTTTTHHHTH & HTTTTHTHHH & 0.18812 & 0.18812\\
\hline
[-0.94809  0.13988  0.25424 -0.13006] & HTTHTHHHTT & THTTHHHTHT & 0.23144 & 0.20043\\
\hline
[-0.66457 -0.47827  0.45341  0.35218] & HTTTTHHHTH & HTTHHTHTTH & 0.29977 & 0.26614\\
\hline
[-0.93392 -0.04759 -0.14279 -0.32426] & THHHTHTHT & TTHHHTHTH & 0.25982 & 0.24801\\
\hline
[-0.06813 -0.20031  0.97526 -0.06406] & TTTTHTTTTH & TTTTHTTTTH & 0.22244 & 0.22244\\
\hline
[-0.52828  0.65335 -0.26856  0.47109] & HTHTT & HTHTT & 0.23627 & 0.23627\\
\hline
[-0.62701  0.42767 -0.1176   0.64041] & HTTHT & HTTHT & 0.22121 & 0.22121\\
\hline
[-0.27418  0.40672 -0.46718  0.73563] & HTTTHTT & HTTTHTT & 0.24486 & 0.24486\\
\hline
[-0.09875  0.75277  0.50256 -0.41354] & TTHTTTTT & TTHTTTTT & 0.28736 & 0.28736\\
\hline
[-0.04894 -0.00402 -0.83205  0.55252] & HTTTTHTT & HTTTTHTT & 0.20474 & 0.20474\\
\hline
[-0.68691  0.36726  0.04274 -0.62566] & TTHTTT & TTHTTT & 0.25131 & 0.25131\\
\hline
[-0.06072  0.76411 -0.12676 -0.62959] & TTTTHTTT & TTTTHTTT & 0.27854 & 0.27854\\
\hline
[-0.62191 -0.0639  -0.76511  0.1541] & TTTTH & TTTTH & 0.19609 & 0.19609\\
\hline
[-0.98674  0.06886 -0.1264   0.07503] & HH & HH & 0.16286 & 0.16286\\
\hline
[-0.86814  0.26898  0.38056  0.17075] & HTTTHTHHHTT & HTHHTHTHTTT & 0.22221 & 0.09319\\
\hline
[-0.2836  -0.03982  0.95045 -0.12098] & HTTTHTHTHTTT & HTTTHTHTHTTT & 0.07442 & 0.07442\\
\hline
[-0.45815 -0.60513 -0.62792  0.17215] & TTTTTHHHTTH & TTTTHTHTHTH & 0.2187 & 0.19569\\
\hline
[-0.60091 -0.54151  0.58106  0.08967] & HTHHTTHTH & HTTHHTHTH & 0.16617 & 0.16617\\
\hline
[-0.3671  -0.15162 -0.40285  0.8246] & HTHTHTHTTTTH & HTHTHTHTTTTH & 0.15013 & 0.15013\\
\hline
[-0.33288  0.42797  0.28725  0.78963] & THTTTH & THTTTH & 0.29693 & 0.29693\\
\hline
[-0.84802 -0.1492   0.02132  0.50808] & HTH & HTH & 0.21022 & 0.21022\\
\hline
[-0.88329 -0.28327 -0.28398  0.2427] & THTHTHHTTH & THTHHHTTTH & 0.21036 & 0.21036\\
\hline
[-0.3926  -0.75829  0.34643 -0.38838] & THTHHHTTTT & TTHTHTTTHH & 0.26302 & 0.22761\\
\hline
[-0.85775  0.2746   0.42074 -0.10883] & HTHTHTHHTT & HTHTHTHHTT & 0.12494 & 0.12494\\
\hline
[-0.27497  0.25412  0.69666  0.61195] & TTHTTTH & TTHTTTH & 0.0623 & 0.0623\\
\hline
[-0.47217  0.0121   0.23258  0.85018] & HTTTH & HTTTH & 0.26015 & 0.26015\\
\hline
[-0.67911 -0.46404 -0.35643  0.4432] & HTTHTHHTTH & HTTHTHHTTH & 0.27136 & 0.27136\\
\hline
\end{tabular}
\end{center}
\end{table*}

\section{Conclusions}
We have shown that the tasks of single-qubit state preparation and gate compilation can be modeled as finite MDPs yielding optimally short gate sequences to prepare states or compile gates up to some desired accuracy. These optimal sequences were found to be comparable with independently calculated shortest gate sequences for the same tasks, often agreeing with them exactly. Additionally, we investigated state preparation in the presence of amplitude damping and dephasing noise channels. We found that an agent can learn information about the noise and yield noise-adapted optimal gate sequences that result in a higher fidelity with the target state. This work therefore provides strong evidence that more complicated quantum programming tasks can also be successfully modeled as MDPs. In scenarios where the state or action spaces grow too large for dynamic programming to be applicable, or where the environment dynamics cannot be accurately learned in the simple manner described above, it is therefore highly promising to apply reinforcement learning to find optimally short circuits for particular tasks. Future work should be directed towards using dynamic programming and reinforcement learning methods for noiseless and noisy state preparation and gate compilation for several coupled qubits. 
%

We provide the required programs for qubit state preparation as open-source software, and we make the corresponding raw data of our results openly accessible~\cite{Orth2022}.

\section{Acknowledgments}
M.S.A. was primarily supported by Rigetti Computing during this work, where he wrote an initial version of this manuscript which was posted on the pre-print repository arXiv. M.S.A. and P.P.O. were supported by the U.S. Department of Energy, Office of Science, National Quantum Information Science Research Centers, Superconducting Quantum Materials and Systems Center (SQMS) under the contract No. DE-AC02-07CH11359. M.S.A. is currently supported under this contract through NASA-DOE interagency agreement SAA2-403602, and by USRA NASA Academic Mission Services under contract No. NNA16BD14C. 

M.S.A. would like to thank Erik Davis and Eric Peterson for valuable insights and useful feedback throughout the development of this work. Previous work with Keri McKiernan, Erik Davis, Chad Rigetti and Nima Alidoust directly inspired this current investigation. Joshua Combes and Marcus da Silva provided early feedback and encouragement to explore this work. P.P.O. acknowledges useful discussions with Derek Brandt. 

\bibliographystyle{apsrev4-2}
\bibliography{References}

%
%
%



\begin{appendix}

\section{Optimal value function}
Here, we prove Eq. \ref{eqn:opt-val-fcn}. For a value function generated by any policy, given that the reward is 1 in the target state $t$ and 0 in every other state
\begin{eqnarray}
V(s) &=& p(t\vert s) \left[ 1 + \gamma V(t) \right] + \sum_{s' \neq t} p(s' \vert s) \gamma V(s') \nonumber \\
&=& p(t \vert s) V(t) + \sum_{s'} p(s' \vert s) \gamma V(s') - p(t\vert s) \gamma V(t) \nonumber \\
&=& p(t \vert s) V(t) (1 - \gamma) + \gamma \sum_{s'} p(s' \vert s) V(s') \nonumber \\
&=& p(t \vert s) V(t) (1 - \gamma) + \gamma \sum_{s'} p(s' \vert s) \times  \nonumber \\
&& \left[ p(t \vert s') V(t) - p(t \vert s') \gamma V(t) + \sum_{s''} p(s'' \vert s') \gamma V(s'') \right] \nonumber \\
&=& p(t\vert s) V(t) (1 - \gamma) + \gamma \sum_{s'} p(t \vert s') p (s' \vert s) V(t) \nonumber \\
&& - \gamma^2 \sum_{s'} p(t \vert s') p(s' \vert s) V(t) + \gamma^2 \sum_{s',s''} p(s''\vert s')p(s' \vert s) V(s'') \nonumber \\
&=& \gamma^{0} (P)_{t,s} V(t) (1 - \gamma) + \gamma (P^2)_{t,s} V(t) (1-\gamma) \nonumber \\
&& + \gamma^2 \sum_{s''} (P^2)_{s'',s} V(s'') \nonumber \\
&=& \sum_{k=0}^{K} \gamma^{k} (P^{k+1})_{t,s} V(t) (1- \gamma) \nonumber \\
&& + \gamma^{K+1} \sum_{s'} (P^{K+1})_{s',s} V(s')
\end{eqnarray}
where in the first equality, we have simply used the fact that the reward is 1 in the target state $t$ and 0 in every other state, in the 4th inequality we have expanded $V(s')$ using the same fact, in the 6th equality we have used the fact that $\sum_{s',s''} p(s'' \vert s') p(s' \vert s) = \sum_{s''} p(s'' \vert s)$ and used the notation $(P^{k})_{s',s} = \sum_{s_1, \dots, s_{k-1}} p(s'\vert s_{k-1}) \dots p(s_{1} \vert s)$, and finally in the last equality we have recursively expanded $V(s'')$, just as in the preceding steps, a total of $K$ times. Noting that we can carry out this recursive expansion arbitrarily many times, in the limit $K \rightarrow \infty$, we find
\begin{equation}
    V(s) = \sum_{k=0}^{\infty} \gamma^{k} (P^{k+1})_{t,s} V(t) (1 - \gamma)
\end{equation}
The above expression is valid for the value function corresponding to an arbitrary policy. Specializing to the optimal policy, for which $V(t) = (1 - \gamma)^{-1}$, we find precisely Eq.~\ref{eqn:opt-val-fcn}.
\end{appendix}

\clearpage

\end{document}